\documentclass[[preprint,twocolumn,preprintnumbers,amsmath,amssymb,superscriptaddress,noshowpacs]{revtex4-1}

\usepackage{graphicx}
\usepackage{dcolumn}
\usepackage{bm}
\usepackage[T2A]{fontenc}
\usepackage[utf8]{inputenc}
\usepackage[english]{babel}
\usepackage{multirow}
\usepackage{xcolor}

\allowdisplaybreaks

\graphicspath{{figs/}}

\begin{document}

\title{The collimated propagation causes \\
	of astrophysical and laboratory jets
}

\author{\firstname{I.}~\surname{Kalashnikov}}
\email{kalasxel@gmail.com}
\affiliation{Keldysh Institute of Applied Mathematics, Russian Academy of Sciences, Moscow, Russia}
\author{\firstname{A.}~\surname{Dodin}}
\affiliation{Sternberg Astronomical Institute, Moscow M.V. Lomonosov State University, Moscow, Russia}

\author{\firstname{I.}~\surname{Ilyichev}}
\affiliation{National Research Center “Kurchatov Institute”, Moscow, Russia}

\author{\firstname{V.}~\surname{Krauz}}
\affiliation{National Research Center “Kurchatov Institute”, Moscow, Russia}

\author{\firstname{V.}~\surname{Chechetkin}}
\affiliation{Keldysh Institute of Applied Mathematics, Russian Academy of Sciences, Moscow, Russia}
\affiliation{National Research Center “Kurchatov Institute”, Moscow, Russia}

%\date{\today}

\begin{abstract}
The use of Z-pinch facilities makes it possible to carry out well-controlled and diagnosable laboratory experiments to study laboratory jets with scaling parameters close to those of the jets from young stars. This makes it possible to observe processes that are inaccessible to astronomical observations. Such experiments are carried out at the PF-3 facility (“plasma focus,” Kurchatov Institute), in which the emitted plasma emission propagates along the drift chamber through the environment at a distance of one meter. The paper presents the results of experiments with helium, in which a successive release of two ejections was observed. An analysis of these results suggests that after the passage of the first supersonic ejection, a region with a low concentration is formed behind it, the so-called vacuum trace, due to which the subsequent ejection practically does not experience environmental resistance and propagates being collimated. The numerical modeling of the propagation of two ejections presented in the paper confirms this point of view. Using scaling laws and appropriate numerical simulations of astrophysical ejections, it is shown that this effect can also be significant for the jets of young stars.
\end{abstract}

\maketitle

\section{Introduction}\label{sec:intro}
In the study of the processes occurring in cosmic space, laboratory modeling begins to play an increasing role. Despite the fact that the characteristic spatial and time scales of laboratory experiments are many orders of magnitude smaller than those of astrophysical ones, they can be scaled to astrophysical ones to the extent that both obey the same equations that do not have their own scale  \cite{Ryutov1999,Ryutov2002}. At the same time, the study of such flows {\it in vitro} makes it possible to measure the available parameters much better than could be done with observations. In addition, laboratory conditions can be changed, thereby studying the response of the system to external influences, which is extremely important for testing the predictions of theoretical models.

To date, many groups around the world are trying, using the laws of scaling, to recreate experimentally
astrophysical conditions during accretion, collimation and subsequent propagation of plasma near young
stars. As a rule, in such experiments, the dynamics of matter is studied on scales of about several centimeters, and the generation of plasma formations usually occurs under conditions close to vacuum. Similar experiments are carried out both on plasma facilities \cite{Lebedev2005,Hsu2005} and on laser ones \cite{Coker2007,Albertazzi2014,Belyaev2018b}.

Interest is focused on laboratory modeling of jet ejections of young stars due to the fact that, as a rule, it is possible to achieve only nonrelativistic velocities in experiments, which are just characteristic of jets of young stars. In this case, a single ejection of a plasma clot is often realized under laboratory conditions. However, according to astronomical observations \cite{Hartigan2001}, at some distance from the young star, jet ejections disintegrate into separate fragments, which at the dawn of interstellar gas observations were named Herbig-Haro objects \cite{Herbig1950}. Therefore, laboratory experiments with single ejections can correspond to space jets
viewed at a considerable distance from the central object.

For laboratory modeling of the jet propagation in the vicinity of a young star, both the spatial scale of the stream and the presence of the ambient environment are essential factors. It is possible to implement these conditions with the PF-3 facility (''plasma focus'', Kurchatov Institute) \cite{Mitrofanov2014,Krauz2018}, which makes it possible to simulate the interaction of astrophysical jets with the surrounding matter in the laboratory. In this case, due to the rather large size of the resulting jets, we can measure not only the thermodynamic and kinematic characteristics, but also, using magnetic probes, the magnitude and structure of magnetic fields.

In our previous paper \cite{Kalashnikov2018}, using numerical simulations, we showed that both in astrophysical and laboratory conditions, after the passage of a supersonic plasma ejection through the environment, a region with low density remains behind it, the so-called vacuum trace. Due to this, subsequent ejections experience much less resistance from the ambient and spread in a more collimated manner—almost all the matter of such an ejection remains within its original radial boundaries. Estimates have shown that the vacuum trace is filled with the surrounding material in at least
$10\text{ $\mu$s}$ in the case of laboratory jets and in $70\text{ yrs}$ in the case of astrophysical jets, i.e., such an effect will be observed if the interval between outflows does not exceed the indicated values. In recent experiments on the PF-3 facility, two ejections were obtained, emitted sequentially from the pinching region. This paper is devoted to the study of this result and its implications for astrophysical applications.

\section{LABORATORY MODELING}\label{sec:exper}

\subsection{PF-3 Facility}\label{subsec:pf3}

\begin{figure*}[ht]
	\centering
	\includegraphics[width=1\linewidth]{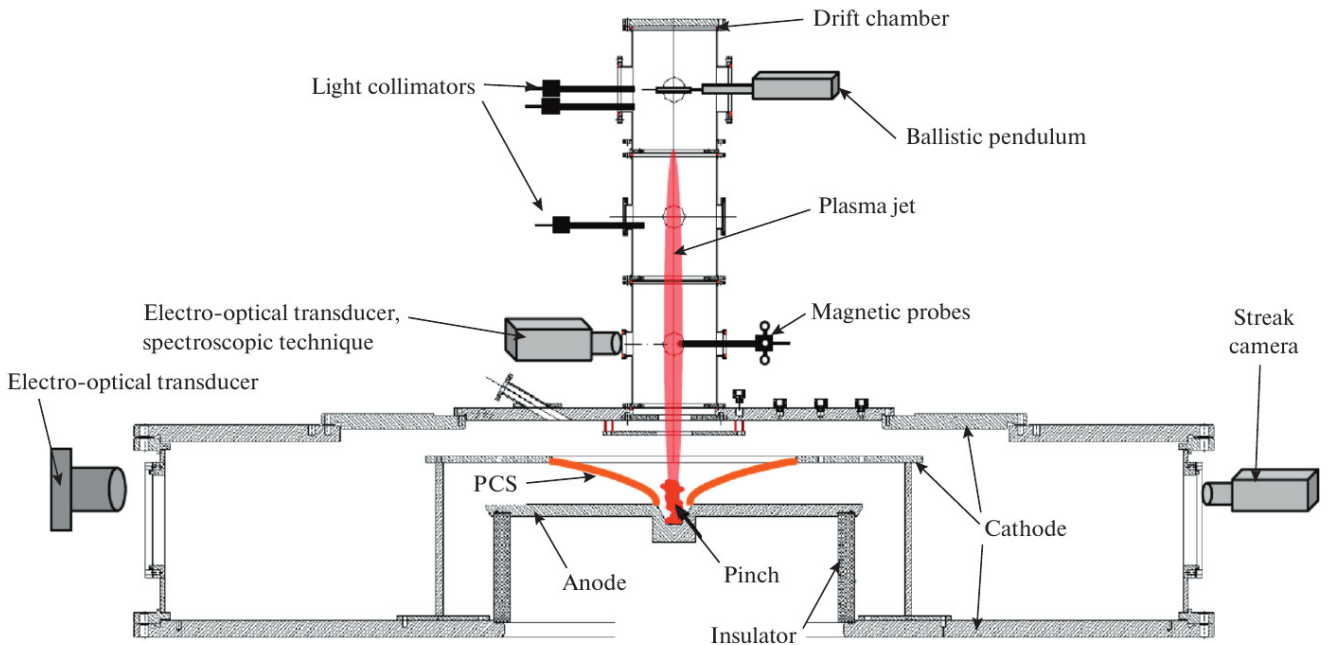}
	\caption{Experiment scheme: PCS -- plasma-current sheath.}
	\label{facility_scheme_p}
\end{figure*}

The PF-3 facility is a plasma focus with a flat geometry of the Filippov-type electrode system \cite{Filippov1962}. A three-section drift chamber is connected to the discharge chamber, which makes it possible to study the dynamics of plasma ejection at a distance of up to  $100 \text{ см}$ from the anode (Fig. \ref{facility_scheme_p}). Each section of the drift chamber is equipped with a set of diagnostic windows designed to measure various parameters at distances of $35$, $65$ и $95 \text{ см}$ (central sections of each window). The PF-3 plasma facility operates in the following way: atmospheric air is preliminary evacuated from the entire facility, then the chambers are filled with some working gas under a pressure of several Torrs. Furthermore, a charged capacitor bank is closed to the cathode and anode, due to which a high voltage appears between the anode and the cathode, which breaks through the working gas. Because of this, a plasma-current sheath (PCS) is formed, which, due to the attraction of parallel currents (the action of the Ampere force), moves to the discharge axis, along which the plasma pinch is formed.

The resulting plasma pinch several centimeters long is characterized by a temperature of the order of $T \simeq 5\cdot 10^2\text{ eV}$ and a concentration of the order of $n\simeq 10^{19}\text{ cm}^{-3}$. At this stage, the current density $10^7 \text{ A/cm}^2$, which leads to the formation of
strong instabilities, the appearance of turbulent resistance, and, as a consequence, an abrupt current cutoff. Thus, the current is disconnected, and the energy stored in the magnetic field of the pinch is transferred to the plasma -- the pinch begins to break and plasma streams are generated, propagating upward along the axis of the drift chamber. The initial stream velocity is $\sim 10^7 \text{ cm/s}$, exceeds the velocity of the PCS motion in the axial direction and is practically independent of the type of working gas.

\begin{figure}[h]
	\centering
	\includegraphics[width=1\linewidth]{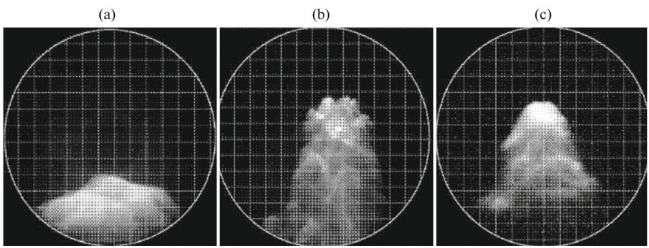}
	\caption{Photos of the plasma stream front according to \cite{beskin2016} at a distance of $z = 35\text{ cm}$ from the anode in a discharge in hydrogen (a) and neon (b), as well as in a discharge in neon at a distance of 65 cm (c). The scale of one cell is  $1\text{ cm}$.}
	\label{jetLabPhoto_p}
\end{figure}

When the already generated stream propagates through the drift chamber, almost all the ejected
plasma remains close to the initial radial boundaries at distances significantly exceeding the transverse dimensions of the stream. This means that the longitudinal velocity of the jet is much higher than the velocity of its transverse expansion. Fig.~\ref{jetLabPhoto_p} shows, as an example, photographs of the stream obtained at different distances in experiments with different gases. It can be seen that the stream structures depend on the chemical composition of the working gas, which is most likely associated with the influence of radiation cooling effects and different sound speed. Shock
waves at the stream front are also clearly visible (analogous to the Herbig–Haro objects). The head part of the stream, even at large distances from the pinching region, has a transverse dimension of several centimeters.

As shown by plasma diagnostics with spectral methods \cite{Ananyev2016}, for a discharge in helium, the background plasma concentration at a distance of  $35\text{ cm}$ is $n_i\approx 2\cdot 10^{16}\text{cm}^{-3}$ (with a temperature $T\approx 1\text{ eV}$). This means that the plasma is $20\%$ ionized. In this case, the plasma stream itself has a concentration $n_i\approx 2\cdot 10^{17}\text{cm}^{-3}$ and a temperature of the order of $T\approx 5\text{ eV}$. Using the magnetoprobe method, it was shown \cite{Mitrofanov2017} that the plasma stream propagates with an already frozen-in magnetic field, the main component of which is a toroidal component with intensity $\sim 10^3\text{ Gs}$, and the poloidal one, at least at some distance from the axis, is an order of magnitude smaller. The studies carried out also made it possible to study the radial distribution of the toroidal field. When $r>r_\text{core}$, the behavior of the field is well described by the dependence $B_\phi\sim 1/r$,
and when $r<r_\text{core}$ it turns out $B_\phi\sim r$ , which will be further used by us for numerical modeling.

\subsection{Results of Experiments with Helium}\label{subsec:experHe}

\begin{figure*}[ht]
	\centering
	\includegraphics[width=1\linewidth]{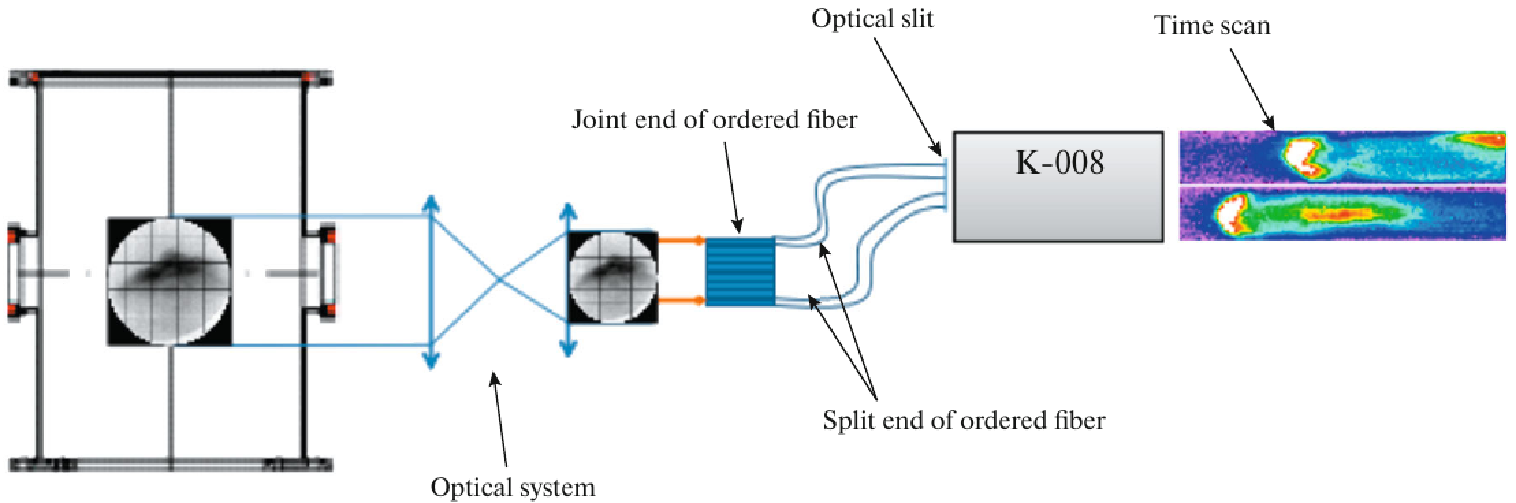}
	\caption{Scheme of measurements with a high-speed streak camera.}
	\label{fig:meas_scheme}
\end{figure*}

\begin{figure*}
	\centering
\includegraphics[width=1\linewidth]{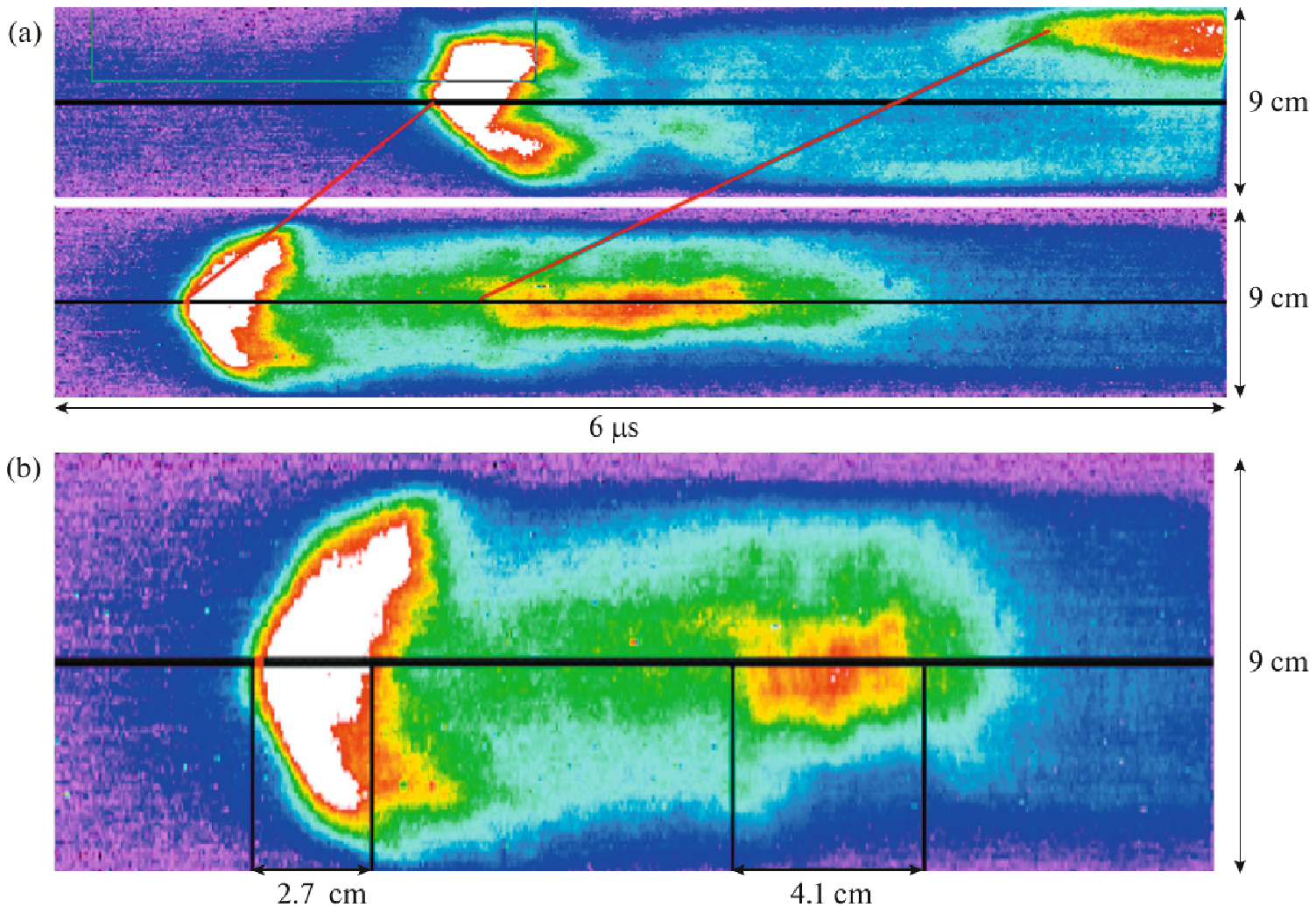}
\caption{Time scan of the plasma stream image at heights of 30.5 and 39.5 cm above the level of the anode (a) and the transformed scan image on a spatial scale (b). The total scan duration is $6\text{ $\mu$s}$ and the image size along the camera diameter is 9 cm. The stream spreads from right to left.}
\label{fig:scan}	
\end{figure*}

As the main diagnostic method in this study, we used the registration of a plasma stream using a high-speed streak camera K-008. The measurement scheme is shown in Fig.~\ref{fig:meas_scheme}. The image of the plasma stream through the diagnostic window is projected by an optical system onto the end face of an ordered fiber, which is a ''sandwich'' of several parallel cores. Thus, each of
the cores transmits to the device input an image from a certain observation area along the height of the camera. At the device input, there is a slit with a width of $0.1\text{ mm}$, which cuts out a narrow region from the object along the diameter of the chamber, which is then unfolded in time by electron-optical methods. An example of such a time scanning is shown in Figs.~\ref{fig:meas_scheme}~and~\ref{fig:scan}a.

First of all, we note that at a height $\approx 30\text{ cm}$, a compact, brightly emitting object with the diameter $\approx 5-6\text{ cm}$ (diameter of the transit chamber is $30\text{ cm}$) and duration of only $\approx 0.5\text{ $\mu$s}$ with a conical profile is observed, apparently due to a shock wave formation.

As noted in Section~\ref{subsec:pf3}, plasma ejections in a plasma focus discharge are formed at the development stage of MHD (magnetohydrodynamic) instabilities, leading to the breakdown of the pinch. The characteristic time of these processes is on the order of hundreds of nanoseconds; therefore, at a considerable distance from the ejection location, the plasma stream is perceived as a single formation with a complex structure (see Fig.~\ref{jetLabPhoto_p}). In this case, it is difficult to identify areas with a vacuum trace. It is almost impossible to organize a re-release in a controlled manner. However, the physics of processes in a plasma focus allows the re-contraction of the current sheath and the formation of a new pinch and, accordingly, the generation of a re-ejection. An example of such a re-ejection is shown in Fig.~\ref{fig:scan}a. In $\approx 1.6\text{ $\mu$s}$, it can be seen that another object of $\approx 1.5\text{ $\mu$s}$ duration appears on the time scan.

Only the duration of the radiation can be measured from the time scan. Longitudinal spatial scales can be roughly estimated using a simple transformation. By the time delay in the appearance of objects on the upper scan relative to the lower one, it is possible to estimate the average stream velocity in the drift region. Thus, for the head object, this velocity is equal to $\sim 7\cdot 10^6 \text{ cm/s}$. For the second bunch, it is two times less. The lower velocity of the second bunch is due to different conditions of stream formation: repeated pinching usually occurs at lower currents, lower pinching velocity, and lower compression ratio. Then, in the first approximation, the length of the bunch will be equal to the product of the bunch velocity by its duration. An example of such a transformation for the lower slit is shown in Fig.~\ref{fig:scan}b.

First of all, attention is drawn to the absence of a shock wave in the second bunch, which is apparently
explained by the motion in the vacuum trace of the first ejection. It should also be noted that the second bunch has a high aspect ratio (>2). Moreover, this is an estimate from below, since the stream velocity in the region of the lower slit is always higher than the due to the deceleration of the stream along the drift length.

\section{SCALING LAWS}\label{sec:scalling}

It was shown for the first time in \cite{Ryutov1999} that two systems described by the equations of an ideal MHD evolve in the same way if they have similar geometry and some scale relations are satisfied. In addition, the conditions for the applicability of ideal MHD are discussed and it is shown that they are satisfied with a large margin both in a number of astrophysical objects and in specially designed experiments with high-power lasers. This makes it possible to carry out laboratory experiments, the results of which can be used for qualitative interpretation of various astrophysical magnetized streams.

The equations of ideal MHD in generally accepted notation have the form:
\begin{align}
&\frac{\partial\rho}{\partial t} = - \operatorname{div} \rho \mathbf{v}, \label{cont} \\
&\frac{\partial\mathbf{v}}{\partial t} + (\mathbf{v},\nabla)\mathbf{v} = - \frac{1}{\rho}\nabla p -\frac{1}{4\pi\rho} [\mathbf{B} \times \operatorname{curl} \mathbf{B}],\\
&\frac{\partial\mathbf{B}}{\partial t} = \operatorname{curl} [\mathbf{v}\times\mathbf{B}],\\
&\operatorname{div} \mathbf{B} = 0,\\ 
&\frac{\partial e}{\partial t} = - \operatorname{div} \left( \mathbf{v} \left(e+p+\frac{\mathbf{B}^2}{8\pi} \right) - \frac{\mathbf{B}(\mathbf{v}\cdot\mathbf{B})}{4\pi}  \right) + S, \label{energ} \\
	& p = (\gamma-1) \left( e - \frac{\rho\mathbf{v}^2}{2} - \frac{\mathbf{B}^2}{8\pi} \right) \label{sost} ,
\end{align}
where $e=\rho\varepsilon +\rho\mathbf{v}^2/{2} + {\mathbf{B}^2}/{8\pi}$ is the total internal
energy and $S$ is the energy sink due to radiation. Knowing the density and pressure, the temperature $T$ can be found from the equation of state of an ideal gas: $ p=\rho R T/M$, where $R$ is the universal gas constant and $M$ is the molar mass. Upon observing $S=0$ and introducing a certain spatial scale $L^*$ in the units in which the lengths in the problem can be measured, we can go over to dimensionless coordinates $\mathbf{r}'=\mathbf{r}/L^*$. Furthermore, we can reach dimensionless of the
quantities appearing in the ideal MHD equations as follows:
\begin{equation}
		\rho'=\rho/\rho^*, \, p'=p/p^*, \, \mathbf{v}' = \sqrt{\rho^*/p^*}\mathbf{v}, \, \mathbf{B}' = \mathbf{B}/\sqrt{p^*},
	\label{transDim_vals}
\end{equation}
where the asterisk denotes the scales of the initial values and the prime denotes the dimensionless functions. From the existing scale values, we can make a combination with the time dimension, in which we can go to dimensionless time $t' = t\sqrt{p_*/\rho^*} / L^*$.

After the indicated transformations, the equation system of ideal MHD will not change, but only dimen-
sionless quantities will appear in it. However, in this case, there are initial conditions which generally can be arbitrary. Combinations of pressure and density scale factors are chosen as scale factors for velocity and magnetic field, but the initial values do not have to be on the same scale. Therefore, for the initial values, we need to write:
\begin{align}
	&\rho|_{t=0} = \rho^* f(\mathbf{r}/L^*), \,\,\, p|_{t=0} = p^* g(\mathbf{r}/L^*),\\ &\mathbf{v}|_{t=0} = v^* \mathbf{h}(\mathbf{r}/L^*), \,\,\, B|_{t=0} = B^* \mathbf{k}(\mathbf{r}/L^*).
	\label{transDim_init}
\end{align}

Considering the above transformations, the following initial conditions are obtained for already dimensionless values:
\begin{align}
	&\rho'|_{t'=0} = f(\mathbf{r}'), \,\,\, p'|_{t'=0} = g(\mathbf{r}'), \label{transDim_initDim1}\\ 
	&\mathbf{v}'|_{t'=0} = Eu \, \mathbf{h}(\mathbf{r}'), \,\,\, B'|_{t'=0} = \sqrt{8\pi/\beta} \, \mathbf{k}(\mathbf{r}'),
	\label{transDim_initDim2}
\end{align}
where $Eu = v^*\sqrt{\rho^*/p^*}$ is Euler’s number and $\beta=8\pi p^*/B^{*2}$ is the ratio of the gas-kinetic pressure of the plasma to the magnetic pressure.

Thus, if any two systems have the same values of $Eu$ and $\beta$, as well as similar initial conditions described by functions $f(\mathbf{r}')$, $g(\mathbf{r}')$, $\mathbf{h}(\mathbf{r}')$, and $\mathbf{k}(\mathbf{r}')$, then their evolution will proceed in the same way. Knowing the behavior of system ''1'', the behavior of system ''2'' can be restored using the specified scaling. For example, the density value of the system ''2'' is expressed through the density value of the system ''1'' as follows:
\begin{equation}
	\rho_2(\mathbf{r},t) = \frac{\rho^*_2}{\rho^*_1} \rho_1    \left(\frac{L^*_2}{L^*_1}\mathbf{r}  , \frac{t^*_2}{t^*_1} t \right).
	\label{scaleRho}
\end{equation} 

With the results of modeling laboratory emissions, it is enough to apply the scaling laws described above and obtain a description for the dynamics of the astrophysical jet propagation. However, these laws do not consider the possibly different radiative characteristics of both systems. Therefore, before asserting that the discovered effect of the vacuum trace formation -- a region with a low concentration left behind a passing supersonic ejection -- takes place in the case of jets from young stars, it is necessary to make sure that the behavior of the system does not qualitatively change
when setting another law of the volumetric cooling rate $S$.

We selected the scale factors, shown in Table~\ref{tab:scal}, based on the data of laboratory experiments and astronomical observations. Densities and pressures were taken in concordance with the state of the background plasma -- helium in a laboratory facility and with the so-called cosmic abundance of elements \cite{Asplund2009} in the astrophysical case.
\begin{table}[h]
	\begin{tabular}{l|c|c|c|c}
		\hline
		Medium  & $L^*$, cm & $\rho^*$, g/cm$^3$ & $p^*$, dyn/cm$^2$ & $t^*$, s \\ 
		\hline
		Laboratory & 5 & $5.4\cdot 10^{-7}$ & $1.3\cdot 10^{5}$ & $10^{-5}$ \\ 
		Astrophysical  & $5\cdot 10^{15}$ & $2\cdot 10^{-21}$ & $4.9\cdot 10^{-10}$ & $10^{10}$ \\ \hline
	\end{tabular}
	\caption{\label{tab:scal} Selected scale parameters corresponding to background plasma}
\end{table}

\section{NUMERICAL SIMULATION}\label{sec:modelling}

\subsection{Laboratory Jets Modeling}\label{subsec:modLab}
Equations (\ref{cont})-(\ref{sost}) were solved using our own program with a Godunov-type numerical scheme in an axisymmetric cylindrical coordinate system using the well-proven HLLD (Harten–Lax–van Leer Discontinuities) solver \cite{Miyoshi2005}. Radiation cooling was considered by calculating the volumetric cooling rate of the plasma as an optically thin body $S= 2k\rho\sigma T^4$, where $\sigma$ is the Stefan–Boltzmann constant and $k$ is the opacity coefficient calculated in the LTE (local thermodynamic equilibrium) approximation using atomic data from TOPbase \cite{Cunto1992}.
	
For a correct comparison of the results of MHD simulation with experiment, the calculation of the
output optical radiation was performed. The radiation transfer equation was solved for cylindrical geometry by the method of short characteristics along beams perpendicular to the propagation axis of the jet. The opacity in lines and continuum was calculated in the LTE approximation using NIST and TOPbase atomic data.

At the initial moment, the ejections were specified as spherical clots of plasma with a radius $R_\text{clot}$ (see Table~\ref{tab:init}), immersed in a background plasma of lower density and temperature. The magnetic field was set only toroidal: it grows linearly from zero to the bunch boundary, and then falls as $r^{-1}$. Although this configuration of the magnetic field corresponds to the measurements carried out on the PF-3 facility, it does not consider the reverse fault currents and the real direction of the magnetic induction vector. But this is quite enough to study the combined effect of the azimuthal field, radiation cooling and ambient environment, as well as the influence of vacuum trace formed by the first ejection on the collimation of the subsequent ejection. The poloidal magnetic field is a more complicated problem due to the need to specify a self-consistent initial configuration of the plasma jet. On the other hand, in the experiment described in Section~\ref{subsec:experHe}, spectral diagnostics of the plasma was not performed, so a significant part of the parameters remained unknown. Thus, due the inconsistent initial configuration of the currents, the goal was not to reproduce the results of the experiment numerically in detail; it was only supposed to qualitatively study the dynamics of the vacuum trace and obtain a similar morphology of the two ejections. Therefore, Table~\ref{tab:init} values may differ slightly from the values that were in the experiment. A longitudinal velocity is imparted to the entire plasma ejection, and it begins to propagate along the axis. Over time $t_\text{launch}$, exactly the same ejection appears in the same place, but with a slightly lower velocity, which apparently took place during laboratory experiments with helium.

\begin{table}[h]
	\begin{tabular}{ll|c|c}
		\hline
		\multicolumn{2}{l|}{Parameter}  & Laboratory & Astrophysical \\ \hline
		\multicolumn{2}{l|}{$R_\text{clot}$, cm} & $2$ & $2\cdot 10^{5}$ \\ %\hline
		\multirow{2}{*}{$n$, cm$^{-3}$}  & Jet  &   $1.8\cdot 10^{17}$  &  3582  \\
		& Background &  $6\cdot 10^{16}$   &   1195  \\ %\hline
		\multirow{2}{*}{$T$, eV} & Jet &  $6$   &  1.52   \\
		& Background &  $1$   &  0.26    \\ %\hline
		\multicolumn{2}{l|}{$B_\phi^\text{max}$, Gs}   &  $7\cdot 10^{3}$   &  $4.3\cdot 10^{-4}$  \\ %\hline
		\multicolumn{2}{l|}{$V_z^{(1)}$, km/s}  &   $55$ & 55.55 \\ %\hline
		\multicolumn{2}{l|}{$V_z^{(2)}$, km/s}  &   $45$ & 45.45 \\ %\hline
		\multicolumn{2}{l|}{$t_\text{launch}$, s} & $5\cdot 10^{-6}$ & $5\cdot 10^{9}$ \\ \hline
	\end{tabular}
	\caption{\label{tab:init} Nonzero initial conditions for simulating laboratory and astrophysical jets.}
\end{table}  

\begin{figure*}[ht!]	
	\begin{minipage}{0.24\linewidth}
		\includegraphics[width=1\linewidth]{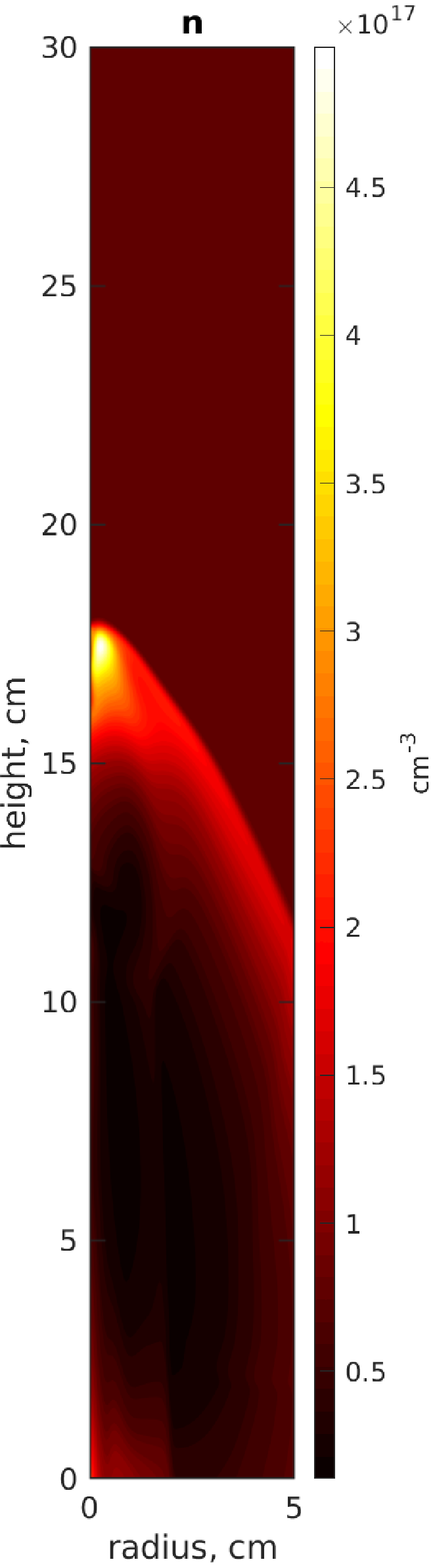} \\ a
	\end{minipage}
	\begin{minipage}{0.24\linewidth}
		\includegraphics[width=1\linewidth]{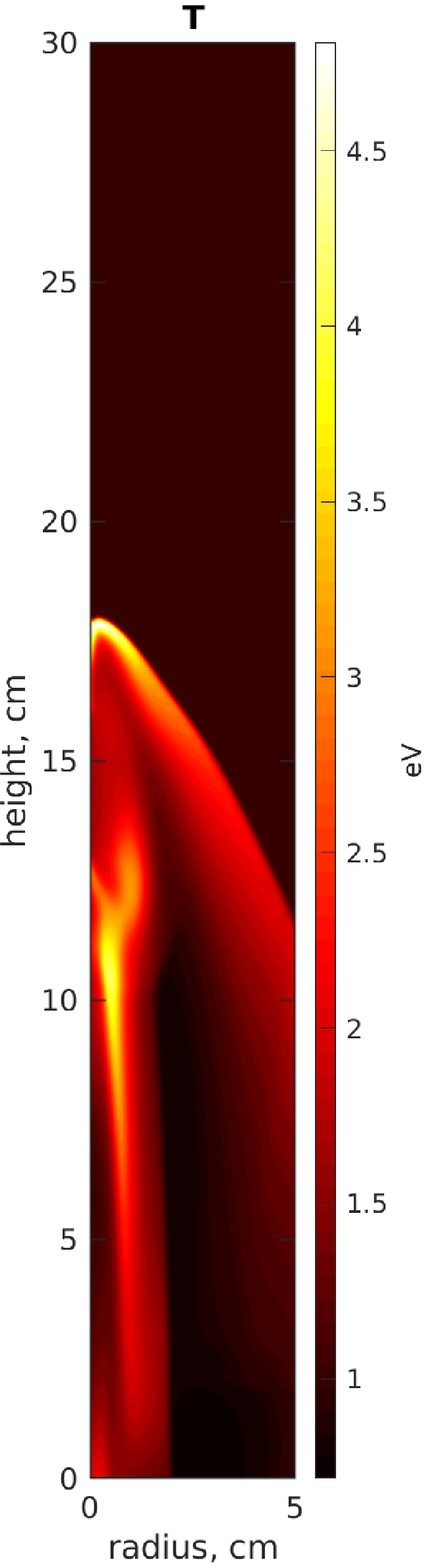} \\ b
	\end{minipage}
	\begin{minipage}{0.24\linewidth}
		\includegraphics[width=1\linewidth]{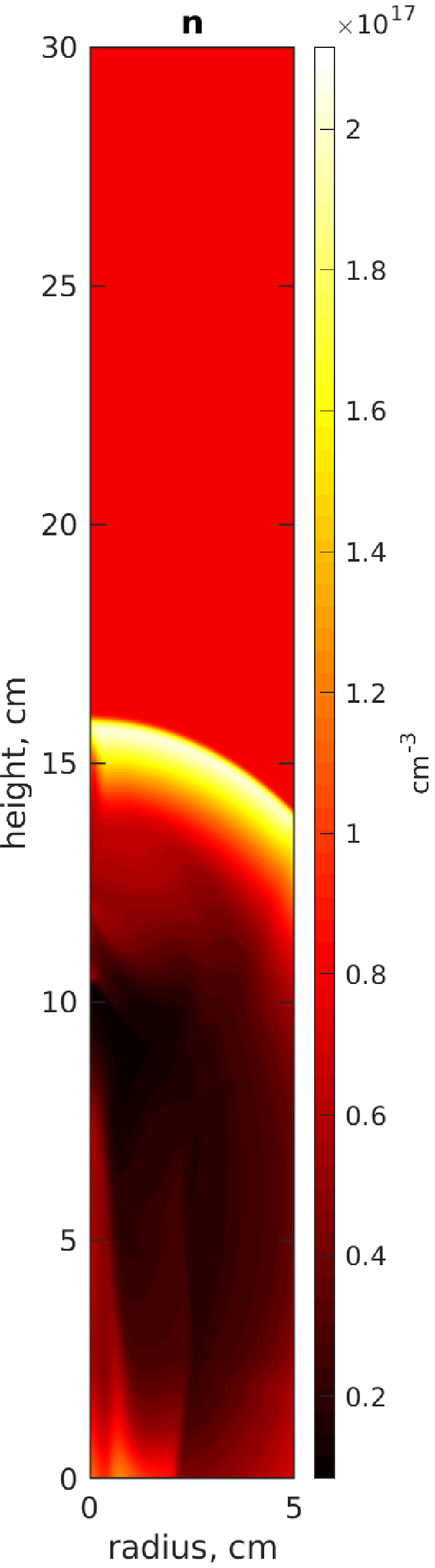} \\ c
	\end{minipage}
	\begin{minipage}{0.24\linewidth}
		\includegraphics[width=1\linewidth]{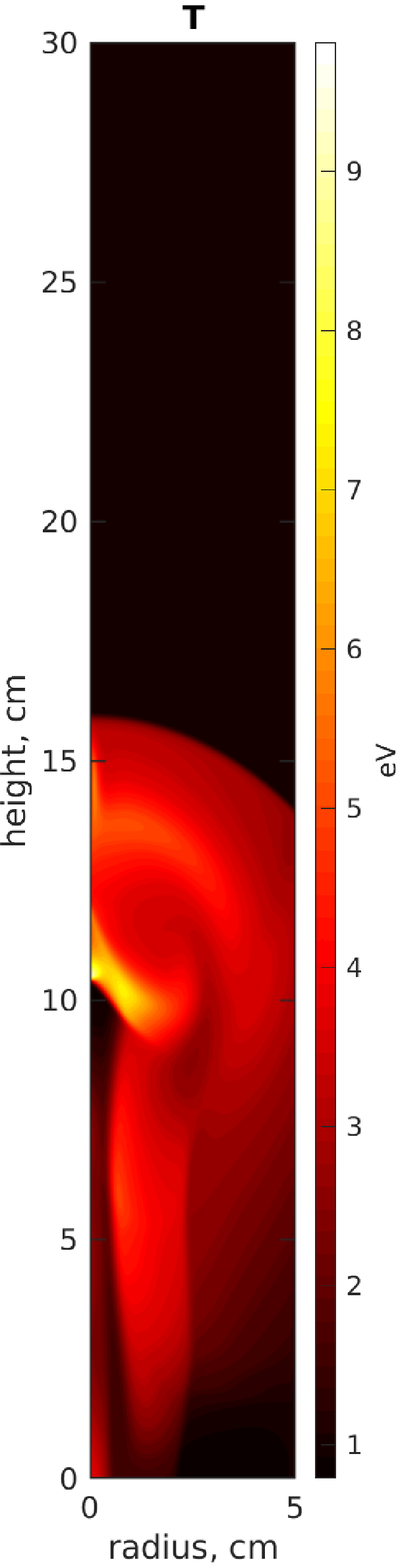} \\ d
	\end{minipage}
\caption{From left to right (a, b, c, d): distribution of concentration (a) and temperature of emitting (b) and non-emitting plasma ((c) and (d), respectively) at the time moment $4.46\text{ $\mu$s}$ in the case of laboratory plasma (working gas is helium).}
\label{fig:lab82}
\end{figure*}

To begin with, let us compare the dynamics of the single jet propagation with and without radiation cooling. This will help to find out how legitimate is the scaling of systems with different term $S$ in expression (\ref{energ}). As we can see in Fig.~\ref{fig:lab82}, the jet dynamics varies considerably. Due to the fact that plasma emits, there is a decrease in temperature and, accordingly, pressure. Therefore, the pressure of the toroidal magnetic field compressing the plasma gradually begins to dominate. Due to this compression in Fig.~\ref{fig:lab82}a, we see the propagation of a compact, relatively cold, plasma clot. While in Fig.~\ref{fig:lab82}c, the initial ejection almost completely disintegrated, forming a mushroom-shaped shock wave. It should be noted that in both cases, a cavity with a low concentration and increased temperature is formed behind the past ejection -- the same vacuum trace. Thus, the specific form of the radiation law $S$ turned out to be not so important for the investigated effect.

\begin{figure*}[ht]
	\begin{minipage}{0.24\linewidth}
		\includegraphics[width=1\linewidth]{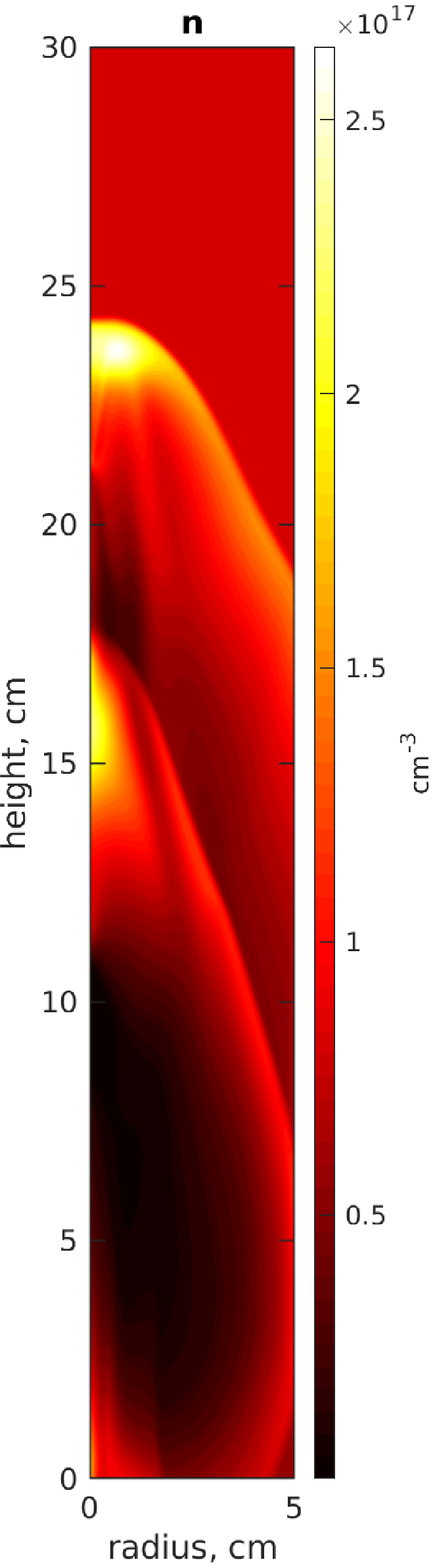} \\ a
	\end{minipage}
	\begin{minipage}{0.24\linewidth}
		\includegraphics[width=1\linewidth]{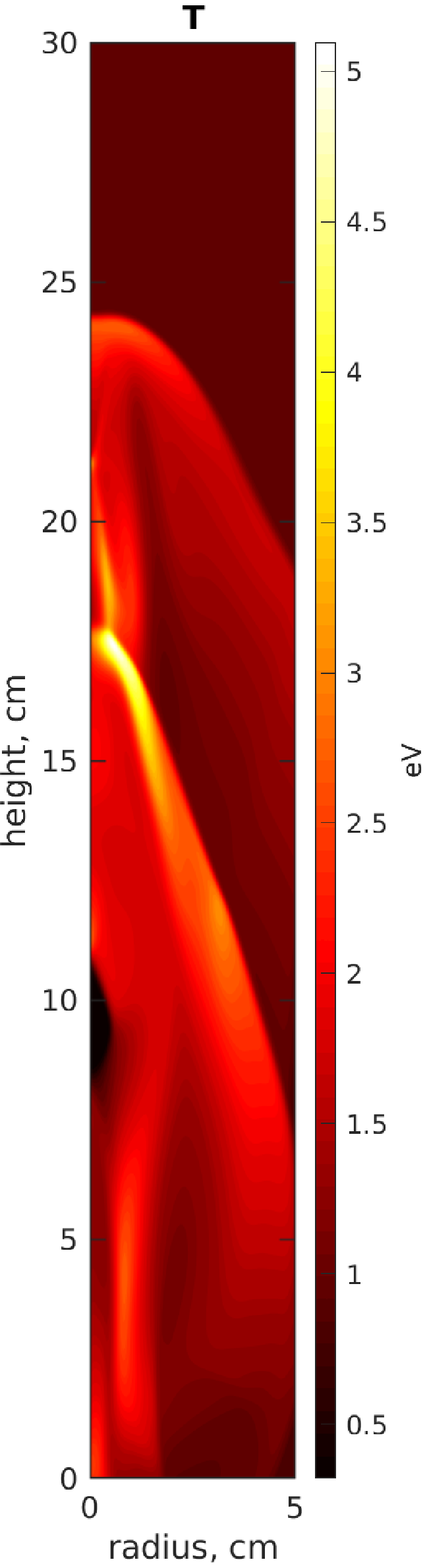} \\ b
	\end{minipage}
	\begin{minipage}{0.24\linewidth}
		\includegraphics[width=1\linewidth]{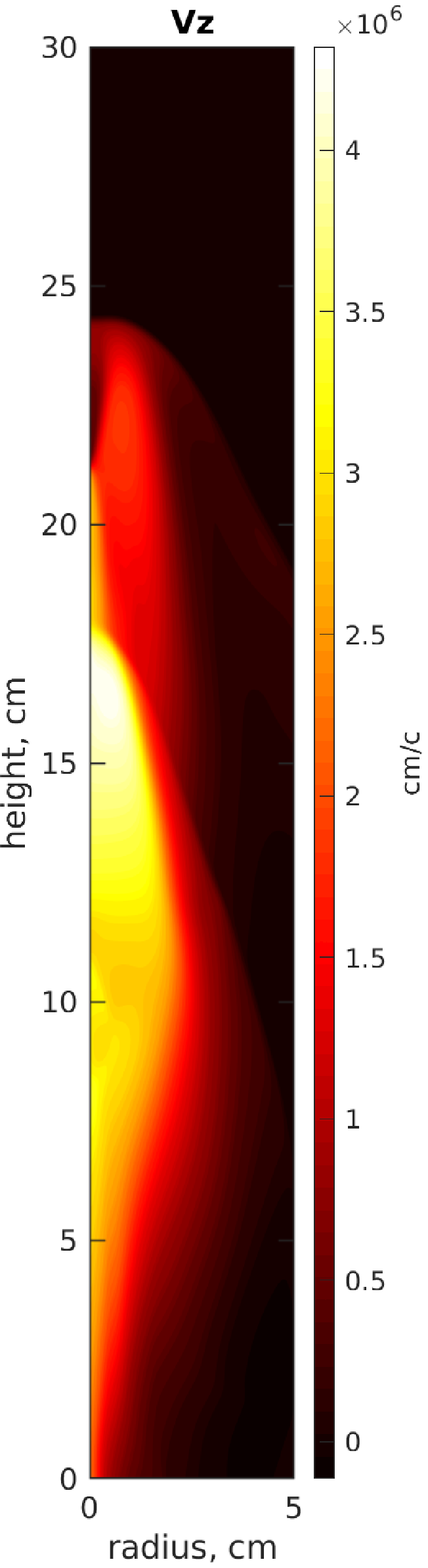} \\ c
	\end{minipage}
	\begin{minipage}{0.24\linewidth}
		\includegraphics[width=1\linewidth]{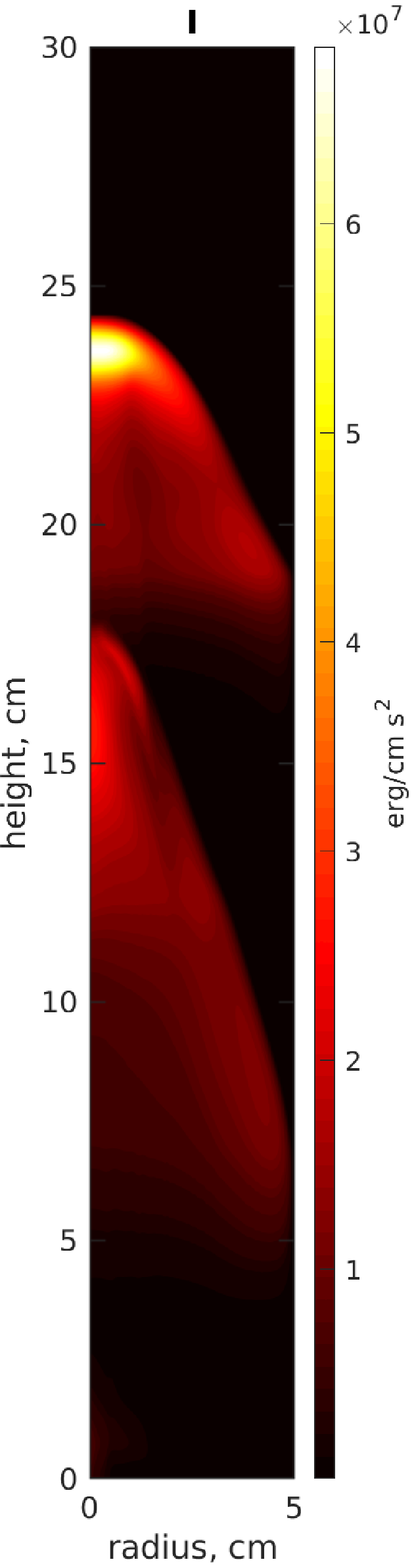} \\ d
	\end{minipage}
\caption{From left to right (a, b, c, d): distribution of concentration (a), temperature (b), longitudinal velocity (c), and radiation intensity (d) at time moment $8.4\text{ $\mu$s}$ in the case of laboratory plasma (working gas is helium).}
\label{fig:lab141}	
\end{figure*}

Let us turn to the case of sequential jet ejection and compare the first burst located at a distance of $16\text{ cm}$ at the time moment $t=4.46\text{ $\mu$s}$ (Figs.~\ref{fig:lab82}a~and~\ref{fig:lab82}b) with the second one, arriving in the same region at the time moment $t=8.4\text{ $\mu$s}$ (Fig.~\ref{fig:lab141}). It can be seen that the second ejection, in comparison with the first one, is more pressed to the axis and forms a shock wave approximately three times less dense. Thus, the ejection following the first one is in the vacuum trace and experiences less resistance from the ambient environment. As we can see in Figs.~\ref{fig:lab82}b~and~\ref{fig:lab141}b, the temperature inside the vacuum trace is several times higher than the temperature of the background plasma. It should be noted that the longitudinal velocity of the matter remaining in the trace is increased (Fig.~\ref{fig:lab141}), which reduces shock compression. Judging by the calculated radiation intensity in the visible range (Fig.~\ref{fig:lab141}d), we were able to satisfactorily reproduce the morphology
of the ejections observed in the experiment (Fig.~\ref{fig:scan}).

Thus, both laboratory experiments described in Section~\ref{subsec:experHe} and the results of numerical simulations for hydrogen, argon, presented in our previous paper \cite{Kalashnikov2018}, and for helium, contained in this section. According to the scaling laws (\ref{transDim_init})-(\ref{scaleRho}) and the results of modeling the single jet propagation with and without radiation cooling, a similar behavior would be expected from ejections whose parameters are close to astrophysical ones. Let us check this by choosing the dimensionless parameters according to Table~\ref{tab:scal}.

\subsection{Astrophysical Jets Modeling}\label{subsec:modAstro}
The simulation procedure described above was performed for astrophysical jets with the initial conditions given in Table~\ref{tab:init}, which correspond to scaling according to laws (\ref{transDim_init})-(\ref{scaleRho}) according to Table~\ref{tab:scal}. However, the volumetric cooling rate of the plasma was calculated as $S= n_e n_i \Lambda$, where $n_e$ and $n_i$ are the volumetric concentrations of electrons and ions, respectively. Also, $\Lambda(T)$ is the cooling function calculated in the CHIANTI package \cite{Dere1997} under the assumption of the standard chemical element abundance in the jet matter \cite{Asplund2009}.

\begin{figure*}[ht]	
	\begin{minipage}{0.24\linewidth}
		\includegraphics[width=1\linewidth]{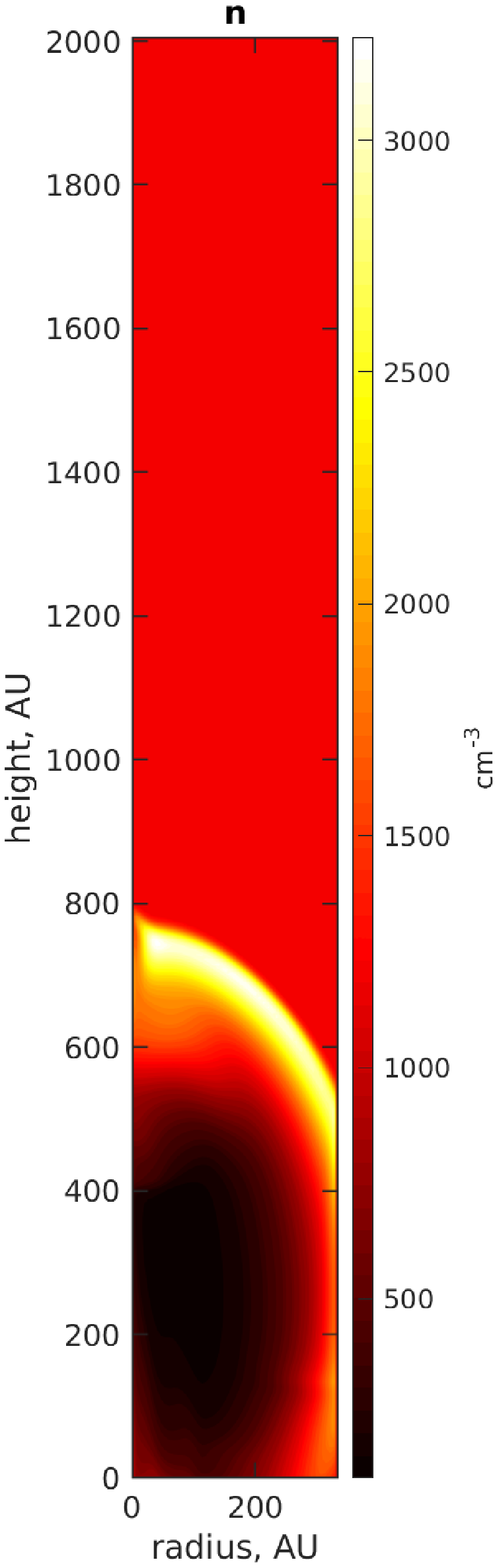} \\ a
	\end{minipage}
	\begin{minipage}{0.24\linewidth}
		\includegraphics[width=1\linewidth]{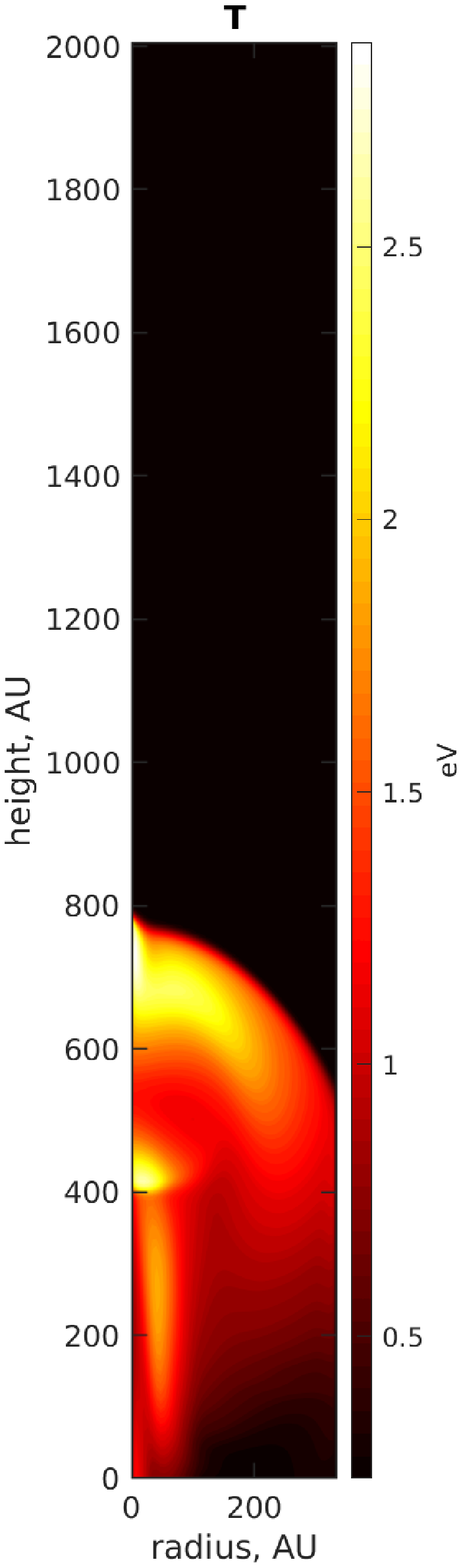} \\ b
	\end{minipage}
	\begin{minipage}{0.24\linewidth}
		\includegraphics[width=1\linewidth]{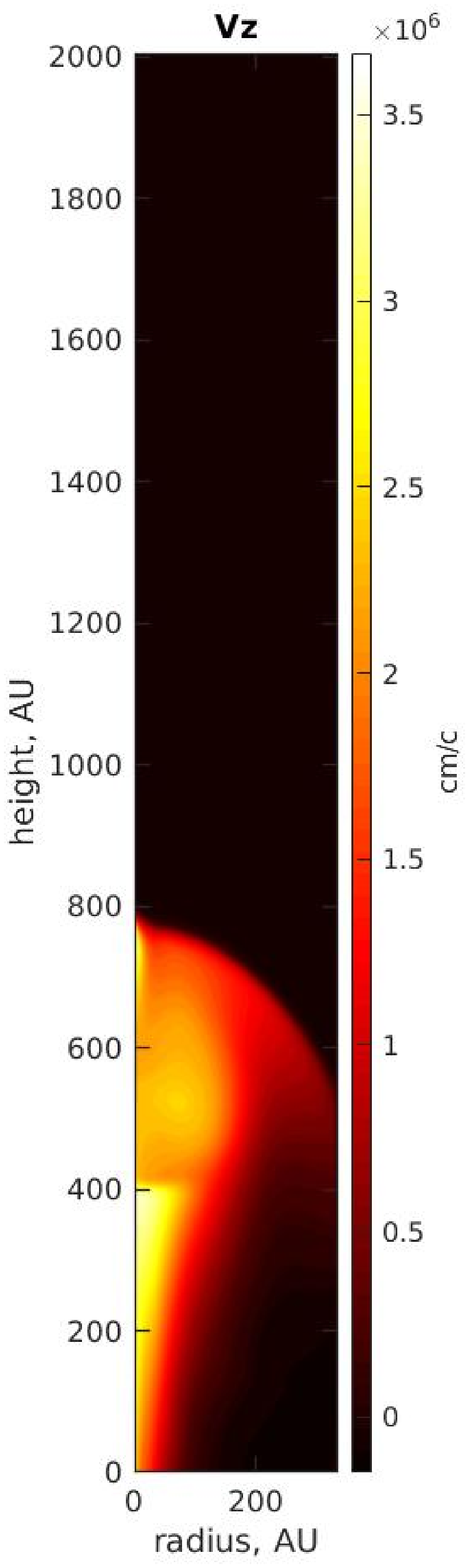} \\ c
	\end{minipage}
	\begin{minipage}{0.24\linewidth}
		\includegraphics[width=1\linewidth]{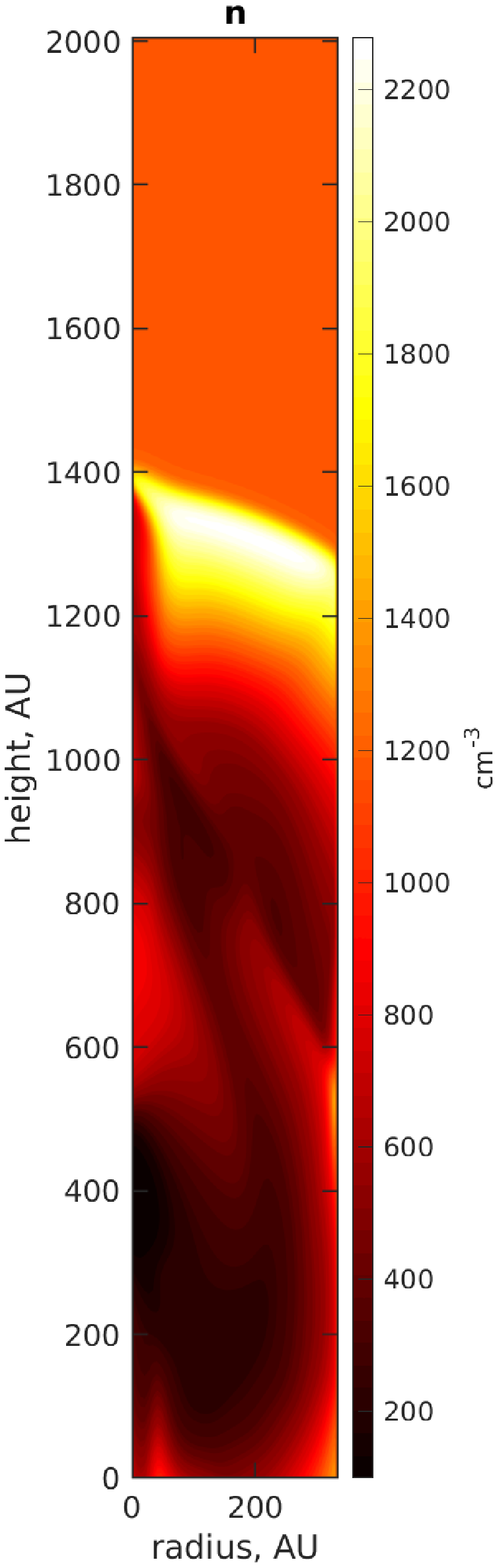} \\ d
	\end{minipage}	
	\caption{ From left to right (a, b, c, d): distribution of concentration (a), temperature (b) and longitudinal velocity (c) at the time moment 78.01 years and concentration at the time moment 268.26 years (d) in the case of astrophysical plasma.}
	\label{fig:astr42p142}	
\end{figure*}

The calculation results for astrophysical ejections by the time $t=78.01\text{ yrs}$ (Figs.~\ref{fig:astr42p142}a~and~\ref{fig:astr42p142}b) indicate that the obtained distributions of plasma density and temperature differ significantly from those for laboratory emitting plasma (Figs.~\ref{fig:lab141}a~and~\ref{fig:lab141}b), and the structure of the bow shock wave is more similar to non-emitting (Figs.~\ref{fig:lab82}~and~\ref{fig:lab82}d). This is obviously due to the poorer emissivity of astrophysical plasma compared to laboratory one. It can be seen that practically all of the material has spread along the mushroom shock wave. Furthermore, we note that a vacuum trace has formed behind the ejection -- a region with a reduced concentration and an increased temperature. As in laboratory conditions, the longitudinal plasma velocity is high in the vacuum trace.

At the time moment $t=268.26\text{ yrs}$, which corresponds to $8.4\text{ $\mu$s}$ for laboratory conditions, both ejections had time to propagate over considerable distances (Fig.~\ref{fig:astr42p142}d). In this case, the second ejection, being in the place of the first one, demonstrates much greater collimation -- it forms an extremely weak shock wave and is strongly pressed to the axis. This is due to the spread of the ejection in an area of low concentration and elevated temperature, which has a significant longitudinal velocity. Thus, the effect of the vacuum trace formation and its influence on the subsequent ejection scales well from laboratory to astrophysical conditions, despite the different laws of radiation cooling.

\section{THE BALANCE OF FORCES ON THE SHOCK-WAVE FRONT}\label{subsec:forces}

\begin{figure}[t]
	\centering
	\includegraphics[width=1\linewidth]{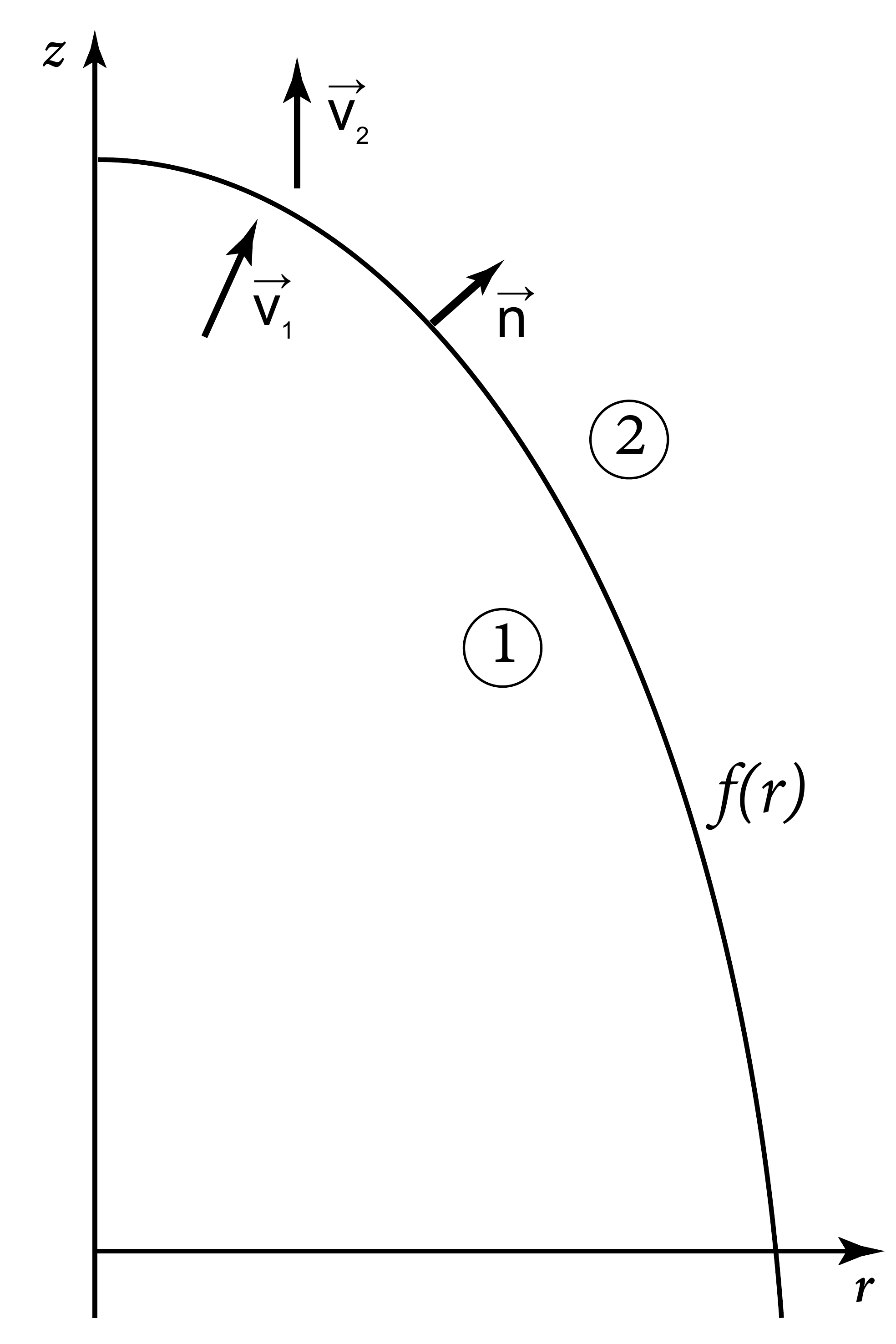}
	\caption{Schematic view of the shock front described by a function $z=f(r)$.	}
	\label{fig:sw_jet}
\end{figure}

Let us analyze the forces acting on the shock-wave front of a jet incident on an ambient environment that has only longitudinal velocity, which is in good agreement with our calculations. For simplicity, we restrict ourselves to the hydrodynamic case only. In the axisymmetric case, the shape of such a front is a rotation shape, described by a certain function $z=f(r)$ (Fig.~\ref{fig:sw_jet}), the normal to which is $\mathbf{n} = \{-f',0,1\}$, where the prime denotes the derivative. Let us denote the
values related to the jet by the subscript ''1'', and to the ambient environment by the subscript ''2''. The velocity of the jet material of the ahead of the shock wave will be denoted as $\mathbf{v_2} = \{u,0,v\}$, the velocity of the surrounding substance as $\mathbf{v_1} = \{0,0,w\}$, and the velocity of the shock front itself will be denoted as $\mathbf{U} = \{C,0,D\}$.

The flow of matter at the discontinuity surface must be continuous; therefore, we have $[\rho(\mathbf{v}-\mathbf{U})\cdot \mathbf{n}] = 0$, where square brackets denote the difference in values on both sides of the shock wave. Hence, we have
\begin{align}
	(\rho_1-\rho_2)(f'C-D) = \rho_2 w + \rho_1(f'u-v). \label{massCont}
\end{align}
Furthermore, the forces with which both gases act on each other must be balanced. Thus, there must be a
continuous flow of impulse through the shock wave: $[\Pi_{ik}n_k]=0$, where $\Pi_{ik} = \rho (v_i-U_i) (v_k-U_k)+p \delta_{ik} $. Taking into account (\ref{massCont}), $r$ and $z$ projections have the form:
\begin{align}
	& (p_1-p_2)f' = \rho_2 u (w+f'C-D), \label{impContR} \\
	& (p_1-p_2) = \rho_2(w-v)(w+f'C-D). \label{impContZ}
\end{align}
From these ratios, it immediately follows:
\begin{align}
	f' = \frac{u}{w-v}. \label{fdef}
\end{align}
Substituting (\ref{fdef}) into (\ref{massCont}),(\ref{impContR}), we
come to the well-known relation \cite{ll6}:
\begin{align}
	\rho_1\rho_2 (u^2 + (v-w)^2) = (\rho_1-\rho_2)(p_1-p_2) \label{velsEoS}.
\end{align}

From (\ref{fdef}) and (\ref{velsEoS}), we obtain the following relation:
\begin{align}
	f' =-\sqrt{1- \frac{\rho_1-\rho_2}{\rho_1\rho_2}\frac{p_1 - p_2}{(v-w)^2}}, \label{fder}
\end{align}
where the minus sign in front of the root was chosen to match the shape of the mushroom shock wave. 

It can be seen from (\ref{fder}) that a decrease in the density or pressure of the ambient environment leads to a decrease in the value at the root, and hence to a steeper behavior of the function $f(r)$. An increase in the longitudinal velocity of the ambient environment $w$ leads to the same effect. Thus, a decrease in density and pressure, as well as an increase in the longitudinal velocity of the external medium, makes the shock wave more transversely compressed, increasing the jet collimation.

\section{CONCLUSIONS}

The simulation results carried out in our previous paper \cite{Kalashnikov2018} predicted the vacuum trace formation following the passage of the first ejection and its critical influence on the collimation of subsequent ones if the working gas in the laboratory facility is argon. Subsequent experiments with helium as the working gas confirmed this effect, which is presented in Section~\ref{subsec:experHe}. Based on numerical simulation for the case of helium plasma, it was possible to recreate the morphology of the ejections observed in experiments and to obtain the spatial distributions of the basic values (Fig.~\ref{fig:lab141}).

Using scaling laws (\ref{transDim_init})-(\ref{scaleRho}), the parameters corresponding to astrophysical ones were selected (Tables~\ref{tab:scal}~and~\ref{tab:init}). The resulting values do not fully correspond to the usual space conditions. Namely, the ratio of the jet density and the ambient environment under astrophysical conditions is much higher than that in the laboratory. In addition, laboratory and astrophysical plasma energy loss due to radiation according to different laws. For parameters close to those observed for object HH 229, we have already performed numerical modeling of successive ejections \cite{Kalashnikov2018}, and the effect associated with the vacuum trace was confirmed. To find out to what extent the effect discovered in the laboratory is applicable to astrophysics, a numerical simulation was performed with the resulting scaling parameters. Its results indicate the vacuum trace formation and the collimation of the plasma ejection following along it (Fig.~\ref{fig:astr42p142}).

The calculations performed with various parameters and simple estimates of the jumps in the values on
the shock wave make one think that the effect of the vacuum trace formation and its influence on subsequent ejections is very universal. Two main characteristics of the vacuum trace can be distinguished, in which the plasma ejections propagating in it remain collimated:
\begin{itemize}
	\item low concentration, with ejections experiencing less resistance;	
	\item significant longitudinal velocity, with relative velocity of the ambient medium and the incident flow decreasing. 
\end{itemize}

Thus, experiments on the PF-3 facility have demonstrated the possibility of reproducing individual
astrophysical effects that occur far from the central object. Thanks to these experiments, it becomes possible to study in the laboratory issues related to the internal structure of astrophysical jets, as well as make informed conclusions about their stability.

The study of the possibility of the vacuum trace formation under astrophysical conditions requires fur-
ther, more accurate theoretical research, considering the reasons for the formation of jets. At least, as shown in this paper, there is reason to believe that the vacuum trace formation in the ejections of young stars is possible.

\smallskip\smallskip\smallskip
%\section*{благодарности}
The authors are grateful to V.S.~Beskin and S.A.~Lamzin for useful discussions, and also to V.V.~Myalton and A.M.~Kharrasov for their help in carrying out the experiments.

The work related to the laboratory facility PF-3 was carried out with the support of the Russian Foundation for Basic Research (project~No.~18-29-21006). The work on numerical modeling was supported by the Russian Scientific Foundation (project~No.~20-11-20165).

\bibliography{lib}

\begin{flushright}
\textit{Translated by E. Seifina}
\end{flushright}

\end{document}